# Transition rate and gravitational wave spectrum from first-order QCD phase transitions

Jingdong Shao[1,*], Hong Mao[2,†] and Mei Huang[3,‡]

[1]*School of Physical Sciences, University of Chinese Academy of Sciences, Beijing 100049, China*
[2]*School of Physics, Hangzhou Normal University, Hangzhou 311121, China*
[3]*School of Nuclear Science and Technology, University of Chinese Academy of Sciences, Beijing 100049, China*

We investigate the gravitational wave spectrum induced by first-order quantum chromodynamics (QCD) phase transitions, e.g., the confinement-deconfinement phase transition of the pure gluon system and the chiral phase transitions in the quark-meson model and Polyakov quark-meson model. The gravitational wave power spectra are sensitive to the phase transition rate $\beta/H$. All QCD models predict a relatively large phase transition rate in the order of $\beta/H \sim 10^4$ at high temperature region, and the produced gravitational waves lie in the peak frequency region of $10^{-4}$–$0.01$ Hz, corresponding to an energy spectrum in the range of $10^{-8}$–$10^{-7}$, which can be detected by LISA and Taiji. If a high baryon density region is generated through Affleck-Dine baryogenesis or other mechanisms, the baryon chemical potential significantly reduces the phase transition rate, which will drop to the order of $\beta/H \sim 10^1$, thus leading to the production of nanohertz gravitational waves. Furthermore, there exists a critical quark chemical potential with zero phase transition rate $\beta/H = 0$, indicating that the false vacuum will not decay, which supports the formation of primordial quark nuggets in the early Universe.

## I. INTRODUCTION

The quantum chromodynamics (QCD) phase transition, as well as the QCD equation of state at finite temperature and baryon density, are essential for exploring the evolution of the early Universe and the structure of compact stars. Recent simulation results from binary-neutron-star (BNS) mergers [1] suggest that a quark-hadron crossover equation of state, featuring a peak in sound velocity, may be favored at high baryon density and low temperature, while the generation of many intriguing observables, including gravitational waves (GWs) [2–5] and primordial black holes [6,7], in the early Universe demands a more attractive strong first-order phase transition.

For QCD with physical quark mass, lattice results indicate a crossover [8] at high temperature around $T_c \sim 160$ MeV at zero or small chemical potential in both the 2-flavor system [9] and 3-flavor system [10]. Moreover, effective QCD models usually support a crossover at high temperature and small or zero chemical potential. Only a few rare cases give QCD first-order phase transitions, e.g., the chiral phase transition in a massless 3-flavor system [11], the deconfinement phase transition in the pure gluon system [11], and the Friedberg-Lee (FL) model [12–14], and the phase transition in a chirality imbalanced system [15,16]. Therefore, in standard cosmology, it is well accepted that the cosmic QCD phase transition from the quark-gluon phase to the hadronic phase would be a crossover, which takes place at about 10 microseconds after the big bang.

For the electroweak phase transition, the standard model shows a crossover at high temperature, and a first-order phase transition can still be realized beyond the standard model [17], which is also a required condition for baryogenesis (BG) [18,19]. A first-order QCD phase transition is normally found in the case of high baryon chemical potentials. The observed baryon asymmetry is tiny, $\eta_B = n_B/s \sim 10^{-9}$, with $n_B$ being the net baryon density and $s$ the entropy density, which seems to indicate that the baryon density in the early universe is small based on adiabatic expansion. However, nonequilibrium processes such as a first-order phase transition can produce a large amount of entropy, and the baryon density in the early universe is not necessarily small [20]. High baryon density can be naturally produced by the Affleck-Dine (AD) baryogenesis mechanism [21–23] in the early universe. The Affleck-Dine baryogenesis mechanism can produce a large baryon asymmetry, $\eta_B \sim 10^{2-3}$, through a scalar field with conserved $B - L$ number, with $B$ the baryon number and $L$ the lepton number in the $SU(5)$ model. It was shown in Ref. [22] that $\eta_B$ cannot exceed $O(1)$ in the early universe. The high baryon density can be diluted by little inflation subsequently [20,24–27] to meet the small baryon asymmetry in the later evolution of the Universe, and the initial $\mu_B/T$ for the little inflation can be as large as $\mu_B/T \sim 100$ [24]. Furthermore, inhomogeneous high baryon density matter might have been distributed in the early universe through first-order electroweak/QCD phase transitions or fluctuations [28–34], forming primordial quark matter. The existence of primordial quark nuggets (PQNs) with high baryon numbers in the early universe has been carefully investigated in many aspects [30,35–37] since it was proposed by Witten in 1984 [29].

[*]Contact author: shaojingdong19@mails.ucas.ac.cn
[†]Contact author: mao@hznu.edu.cn
[‡]Contact author: huangmei@ucas.ac.cn

GWs from cosmic phase transitions have become a heated topic recently, especially following observations of evidence of stochastic GWs in the nanohertz band [38–41] from several independent pulsar timing array (PTA) collaborations, including the North American Nanohertz Observatory for Gravitational Waves (NANOGrav) [42,43], the European and Chinese PTA (EPTA and CPTA) [44–48], and the Parkes PTA (PPTA) [49]. The nanohertz GWs are possibly generated by the orbiting or merger of supermassive black hole (with mass $10^5$–$10^{10}M_\odot$) binaries [50], or by cosmic phase transitions in the electroweak [51] or QCD epoch [41,52].

Regarding GW spectra from first-order phase transitions, the transition rate $\beta/H$, which measures the inverse duration of cosmic phase transitions, is a pivotal parameter that dictates the peak frequency of GW spectra. However, to generate nanohertz GWs, $\beta/H$ is strictly bounded by various sources, e.g., $\beta/H < 15$ for QCD phase transitions [52], $\beta/H \sim 10$ for cosmic strings [53], and $\beta/H < 12$ for secondary GWs from primordial black holes formed during the first-order phase transition [6]. However, in many references, $\beta/H$ is simply adapted as a free parameter just to yield nanohertz GWs, e.g., $\beta/H$ around 1–10 in the pure gluon system [54,55]. In fact, this strict constraint $\beta/H \sim 10$ is equivalent to a slow phase transition, which can result in bulky true vacuum bubbles before collisions and more inhomogeneous, instead of small dense homogeneous, bubbles in a typical fast phase transition. The slow phase transition consequently brings GWs with lower frequencies and higher peaks, along with other intriguing observables such as the formation of primordial black holes [6,7]. Such a slow phase transition usually demands strong supercooling before nucleation, which can be manually or automatically attributed to phase transitions by specific mechanisms. For instance, electroweak phase transitions with extreme strong supercooling are widely tried to generate nanohertz GWs [51,56–59], but it is difficult to defer electroweak phase transitions until the QCD epoch (100 MeV) [56]. Alternate mechanisms such as a QCD-triggered electroweak phase transition after a heat inflation [60] are also proposed to obtain small $\beta/H$. However, compared with achievable small transition rates and strong supercooling in electroweak phase transitions with special mechanisms, extreme supercooling mechanisms do not work well in QCD phase transitions due to constraints of modern cosmology like Big Bang nucleosynthesis (BBN), and typical values of $\beta/H$ calculated are large. Typical first-order QCD phase transitions at high temperatures in low-energy effective QCD-like theories and holographic QCD models feature larger transition rates $\beta/H \sim 10^{4-5}$ [61–65], and corresponding GW spectra lie in the range of Taiji and LISA with peak frequencies typically $10^{-4}$–$10^{-2}$ Hz and peak energy densities $10^{-8}$–$10^{-7}$. For example, $\beta/H$ is around $10^5$ in the deconfinement phase transition for $SU(N_c)$ gauge theory described by an effective potential with a nonperturbative gluon mass [65]. Besides, the sphaleron transition-induced chirality-imbalanced system experiences a first-order phase transition [15] with a transition rate $\beta/H \sim 10^4$ [16]. For the first-order chiral phase transition, holographic QCD calculations also give transition rates around $10^{4-5}$ [61,62], in agreement with results from low-energy effective chiral models of QCD-like theories [63,64].

In this work, we explore first-order QCD phase transitions in different effective QCD models, including the deconfinement phase transition in the pure gluon system and the Friedberg-Lee model, the chiral phase transition in the quark-meson and the Polyakov quark-meson model. The phase transition strength parameter $\alpha$, the phase transition rate $\beta/H$ and the corresponding GW spectra are calculated accordingly. We check whether it is possible to produce nanohertz GWs from first-order QCD phase transitions, particularly if high baryon density QCD matter can be generated in the early Universe. We will show that the baryon chemical potential can largely reduce the phase transition rate $\beta/H$, and there exists a narrow window of high baryon chemical potential with a transition rate $\beta/H \sim 10^1$ to produce nanohertz GWs. There exists a critical quark chemical potential with zero phase transition rate $\beta/H = 0$, which indicates the phase transition can barely start. Small $\beta/H$ means slow phase transitions and zero $\beta/H$ implies that the phase transition from quark matter to hadronic matter will never complete. It is possible that primordial "quarklet" or primordial quark nuggets existed in the early Universe and can survive for cosmic-scale times.

The article is organized as follows, in Sec. II we briefly introduce the process to calculate GWs from QCDPTs, in Secs. III–V we calculate GWs in the pure gluon system, Friedberg-Lee model, QM model and PQM model with different chemical potential $\mu$ respectively, in Sec. VI we give the summary and discussion about different models.

## II. GW SPECTRA FROM FIRST-ORDER PHASE TRANSITION AT FINITE BARYON DENSITY

Cosmic first-order phase transitions are achieved via nucleation of true vacuum bubbles. When the Universe cools down, some parts of the Universe jump to the true vacuum with lower energy density from the false vacuum and form bubbles, inside of which is the true vacuum, while outside of which remains the false vacuum. The released latent heat is transferred into the energy of bubble walls and, driven by the difference in pressure between the true and false vacuum, these true bubbles expand, continually engulfing the false vacuum part. Once these bubbles collide with each other and pass kinetic energy to the surrounding media, GWs can be generated from three main sources: the collisions of bubbles, the sound waves, and the magneto-hydrodynamic (MHD) turbulence [66].

The nucleation rate of true vacuum bubbles [67–69]

$$\Gamma(t) = A e^{-S_4(t)} \tag{1}$$

is mainly determined by the Euclidean action $S_4$ of an $O(4)$-symmetric bounce solution between the true and false vacuum. While at high temperature $T$, bubbles have an $O(3)$-symmetry instead of $O(4)$, and $S_4$ reduces to $\frac{S_3}{T}$. The coefficient $A$ at high temperature is then [70]

$$A(T) = T^4 \left(\frac{S_3}{2\pi T}\right)^{\frac{3}{2}}. \tag{2}$$

Specifically, for a scalar field $\Phi$ with potential $V(\Phi)$, the bounce action is

$$S_3 = \int d^3r \left(\frac{1}{2}(\nabla\Phi)^2 + V\right) \tag{3}$$

and the configuration of $\Phi(r)$ is given by the $O(3)$-symmetric equation of motion

$$\frac{d^2\Phi}{dr^2} + \frac{2}{r}\frac{d\Phi}{dr} = \frac{dV}{d\Phi}. \tag{4}$$

with boundary conditions $\frac{d\Phi}{dr}|_{r=0} = 0$ and $\Phi(\infty)$ remaining the false vacuum. At high temperature, $V$ is the grand potential $\Omega$ in the finite temperature field theory.

The temperature at which bubbles start to emerge and the nucleation rate catches the expansion rate of the Universe is defined as the nucleation temperature $T_n$. In the QCD epoch, $T_n$ can be quickly estimated by $S_3/T \sim 180$. More precisely, one bubble per Hubble volume per Hubble time is expected at $T_n$ [68,69]

$$\Gamma(t)/H^4 \sim 1, \tag{5}$$

where $H$ is the Hubble parameter given by the Friedmann equation

$$H = \sqrt{\frac{\rho}{3m_p^2}} \tag{6}$$

with the reduced Planck Mass $m_p = 2.435 \times 10^{18}$ GeV and the energy density $\rho$. Approximately, $T_n$ is also the temperature of the thermal bath without strong reheating, thus the transition rate $\beta$ is defined as

$$\frac{\beta}{H} = T_n \frac{d\left(\frac{S_3}{T}\right)}{dT}\bigg|_{T_n}. \tag{7}$$

Alternatively, the nucleation condition can be made through the percolation temperature $T_p$, at which the true vacuum bubbles occupy 30% of the volume. In cases below where $\beta/H \gg 1$, the false vacuum decays so rapidly that the temperature barely changes during the phase transition, thus $T_p \approx T_n$ is appropriately adapted.

Another parameter $\alpha$ that measures the relative transition strength can be written as the ratio of the latent heat released to the background radiation energy density $\rho_r$ [68,70]

$$\alpha = \frac{-\Delta\rho + 3\Delta p}{4\rho_r} = \frac{1}{\rho_r}\left(\Delta p - \frac{T}{4}\frac{\partial \Delta p}{\partial T}\bigg|_{T_p} - \frac{\mu}{4}\frac{\partial \Delta p}{\partial \mu}\bigg|_{T_p}\right), \tag{8}$$

where $\Delta$ means the difference between the true and false vacuum. Here the latent heat between the true and false vacuum is expressed as $\Delta\theta/4$ in terms of the trace of the energy-momentum tensor $\theta = -\rho + 3p$ with pressure $p = -\Omega$ and energy density $\rho = -p + T\frac{\partial p}{\partial T} + \mu\frac{\partial p}{\partial \mu}$. Meanwhile, $\rho_r = \frac{\pi^2 g T^4}{30}$ ($g$ is the number of relativistic degrees of freedom) as in the bag model is not accurate because the thermal background is not the perfect ideal gas, especially when the chemical potential $\mu$ is not negligible compared with temperature $\mu \geq T$. Instead, the thermal energy density is the energy density of the false vacuum

$$\rho_r = -(p - p_{\text{vac}}) + T_n \frac{\partial p}{\partial T}\bigg|_{T_n} + \mu \frac{\partial p}{\partial \mu}, \tag{9}$$

here $p_{\text{vac}}$ is the vacuum pressure at $T = \mu = 0$ and must be deducted. For dense matter $\mu \gtrsim T$, energy density is mostly contributed by the third term due to the high number density $n = \frac{\partial p}{\partial \mu}$.

The dynamics of true vacuum bubbles after nucleation is model dependent and numerical relations between $v_w$ and other parameters can be extracted from different approaches. The bubble wall velocity $v_w$ can be given by the Boltzmann equation in a specific model [71,72], by numerical simulations [73] or by holographic methods [61,74]. Typically $v_w$ becomes relativistic soon after nucleation and moderate velocity (e.g., $v_w > 0.3$) makes little differences, thus here we use a good approximation for strong phase transitions [43,75,76]

$$v_w = v_J = \frac{\sqrt{1/3} + \sqrt{\alpha^2 + 2\alpha/3}}{1 + \alpha}, \quad (10)$$

where $v_J$ is the velocity for Jouguet detonation. Although small $v_w$ also exists in strong phase transitions due to model dependence and influences the shapes of GW spectra, as we will see below, the GW spectra with large $\beta/H$ are not detectable with relativistic $v_w$, let alone small ones. $v_w$ does not change physical facts below, especially the survival of dense QCD matter.

Furthermore, two efficiency factors $\kappa_v$ and $\kappa_{tb}$, the fraction of the vacuum energy converted into the kinetic energy of the background media and the MHD turbulence respectively, can be analytically fitted according to $\alpha$ and $v_w$ [76–78]. Using numerical fits in Ref. [76], $\kappa_v$ can be expressed under different situations. With definitions

$$\kappa_A = v_w^{6/5} \frac{6.9\alpha}{1.36 - 0.037\sqrt{\alpha} + \alpha},$$

$$\kappa_B = \frac{\alpha^{2/5}}{0.017 + (0.997 + \alpha)^{2/5}},$$

$$\kappa_C = \frac{\sqrt{\alpha}}{0.135 + \sqrt{0.98 + \alpha}},$$

$$\kappa_D = \frac{\alpha}{0.73 + 0.083\sqrt{\alpha} + \alpha},$$

$$\delta\kappa = -0.9 \ln\left(\frac{\sqrt{\alpha}}{1 + \sqrt{\alpha}}\right),$$

the efficiency factor $\kappa_v$ is

$$\kappa_v = \frac{v_s^{11/5} \kappa_A \kappa_B}{(v_s^{11/5} - v_w^{11/5})\kappa_B + v_w v_s^{6/5} \kappa_A}$$

for subsonic deflagrations $v_w < v_s$,

$$\kappa_v = \kappa_B + (v_w - v_s)\delta\kappa + \frac{(v_w - v_s)^3}{(v_J - v_s)^3}(\kappa_C - \kappa_B - (v_J - v_s)\delta\kappa)$$

for supersonic deflagrations $v_s < v_w < v_J$ and

$$\kappa = \frac{(v_J - 1)^3 (v_J/v_w)^{5/2} \kappa_C \kappa_D}{((v_J - 1)^3 - (v_w - 1)^3)v_J^{5/2} \kappa_C + (v_w - 1)^3 \kappa_D} \quad (11)$$

for detonations $v_w \geq v_J$ as Eq. (10) indicates. The other efficiency factors $\kappa_{tb}$ is taken as $\kappa_{tb} = 0.05\kappa_v$ according to results from numerical simulations $\frac{\kappa_{tb}}{\kappa_v} \sim 0.05$–$0.1$ [77,79].

In terms of the parameters above, the numerical results of GW spectra from sound waves and MHD turbulence are respectively [70,77]

$$h^2 \Omega_{sw} = 2.65 \times 10^{-6} \left(\frac{H}{\beta}\right) \left(\frac{\kappa_v \alpha}{1 + \alpha}\right)^2 \left(\frac{100}{g}\right)^{\frac{1}{3}} v_w S_{sw}(f) \quad (12)$$

and

$$h^2 \Omega_{tb} = 3.35 \times 10^{-4} \left(\frac{H}{\beta}\right) \left(\frac{\kappa_{tb}\alpha}{1 + \alpha}\right)^2 \left(\frac{100}{g}\right)^{\frac{1}{3}} v_w S_{tb}(f), \quad (13)$$

with

$$S_{sw}(f) = \left(\frac{f}{f_{sw}}\right)^3 \left(\frac{7}{4 + 3(\frac{f}{f_{sw}})^2}\right)^{\frac{7}{2}}, \quad (14)$$

$$S_{tb}(f) = \left(\frac{f}{f_{tb}}\right)^3 \left(1 + \frac{f}{f_{tb}}\right)^{-\frac{11}{3}} \left(1 + \frac{8\pi f}{h}\right)^{-1}. \quad (15)$$

The peak frequencies of the two sources are

$$f_{sw} = 1.9 \times 10^{-5} \frac{1}{v_w} \frac{\beta}{H} \frac{T_n}{100 \text{ GeV}} \left(\frac{g}{100}\right)^{\frac{1}{6}} \text{ Hz} \quad (16)$$

and

$$f_{tb} = 2.7 \times 10^{-5} \frac{1}{v_w} \frac{\beta}{H} \frac{T_n}{100 \text{ GeV}} \left(\frac{g}{100}\right)^{\frac{1}{6}} \text{ Hz} \quad (17)$$

with

$$h = 1.65 \times 10^{-6} \frac{T_n}{100 \text{ GeV}} \left(\frac{g}{100}\right)^{\frac{1}{6}} \text{ Hz} \quad (18)$$

the Hubble rate.

Similarly, the other part of GWs from bubble collisions is [77]

$$h^2 \Omega_{env} = 1.67 \times 10^{-5} \left(\frac{H}{\beta}\right)^2$$
$$\times \left(\frac{\kappa_\phi \alpha}{1 + \alpha}\right)^2 \left(\frac{100}{g}\right)^{\frac{1}{3}} \left(\frac{0.11 v_w^3}{0.42 + v_w^2}\right) S_{env}(f),$$

with

$$S_{env}(f) = \frac{3.8(f/f_{env})^{2.8}}{1 + 2.8(f/f_{env})^{3.8}}$$

and

$$f_{env} = 1.65 \times 10^{-5} \frac{0.62}{1.8 - 0.1 v_w + v_w^2} \frac{T_n}{100 \text{ GeV}} \left(\frac{g}{100}\right)^{\frac{1}{6}} \text{ Hz}.$$

Here $\kappa_\phi$ denotes the fraction of vacuum energy converted into the gradient energy of the scalar field. However, the energy in the scalar field is negligibly small for relativistic bubbles [77] and GWs from the former two sources mainly contribute to the total energy density spectrum, i.e.,

$$Eh^2\Omega = h^2\Omega_{sw} + h^2\Omega_{tb}. \quad (19)$$

## III. THE DECONFINEMENT PHASE TRANSITION OF THE PURE GLUON SYSTEM FROM LATTICE QCD

Lattice QCD calculation shows that the deconfinement phase transition of the pure gluon system is of first-order [80]. The pure gluon system and the corresponding deconfinement phase transition can be described by the Polyakov loop. The thermodynamic Wilson line takes the form of

$$\mathbf{L}(\vec{x}) = P \exp\left(ig \int_0^{1/T} A_4(\vec{x},t)dt\right) \quad (20)$$

and the Polyakov loop is

$$\Phi = \frac{1}{N_c}\mathrm{Tr}(\mathbf{L}(\vec{x})) \quad (21)$$

with $N_c = 3$. Given $SU(3)$ [see Ref. [81] for cases of $SU(N)$], the effective potential can be written as

$$\mathcal{U}(\Phi,\bar{\Phi},T)/T^4 = a(\Phi\bar{\Phi})^2 - \frac{b}{2}(\Phi^3 + \bar{\Phi}^3) + c\Phi\bar{\Phi}, \quad (22)$$

where $a$, $b$ and $c$ are undetermined coefficients. Following Refs. [82,83], $a$ and $b$ are constants while $c$ is a function of temperature, but only two terms are necessarily included here

$$c(T) = c_2 T_0^2/T^2 + c_4. \quad (23)$$

For the pure gluon system, $T_0 = T_c = 276$ MeV. These coefficients listed in Table I are fitted by lattice data [80] and comply with $\lim_{T\to\infty}\Phi = 1$. Meanwhile in the high temperature limit the pressure $p = -\mathcal{U}$ should approach that of the ideal gas $p = \frac{\pi^2}{90}gT^4$ [83], where $g = 16$ is the degree of freedom of gluons. As calculated, the function $c(T)$ does not change the sign as the temperature changes, the potential barrier always exists at $T \geq T_c$ and disappears at small supercooling, while in some Refs. [81,83,84] $c(T)$ changes the sign at $T > T_c$ and simultaneously the potential barrier appears, but the phase transition is of first-order in all cases.

In the absence of quarks, the gap equation as well as the equation of motion is

$$\frac{\partial\mathcal{U}(\Phi,\bar{\Phi},T)}{\partial\Phi} = \frac{\partial\mathcal{U}(\Phi,\bar{\Phi},T)}{\partial\bar{\Phi}} = 0 \quad (24)$$

with $\Phi = \bar{\Phi}$ at zero chemical potential. After simplification, the gap equation

$$4a\Phi^3 - 3b\Phi^2 - 2c\Phi = 0 \quad (25)$$

TABLE I. Parameters of the potential in the pure gluon system fitted by lattice data.

| $a$ | $b$ | $c_2$ | $c_4$ |
|---|---|---|---|
| 7.69 | 12.0 | 2.06 | 2.62 |

TABLE II. Parameters $T_n$, $\beta/H$, and $\alpha$ in the pure gluon system.

| $T_n$/MeV | $\mu$ | $\beta/H$ | $\alpha$ |
|---|---|---|---|
| 270.55 | 0 | 17742 | 0.298 |

is simple and can be solved analytically

$$\Phi = 0; \quad \Phi = \frac{3b \pm \sqrt{9b^2 - 32ac}}{8a}. \quad (26)$$

The first solution is the confinement phase available at any temperature, but the latter two solutions are valid only when $9b^2 - 32ac > 0$ and this inequality gives $\frac{T}{T_c} > 0.8826$, which means the supercooling has an upper limit and the potential barrier vanishes at this limit. The actual supercooling at the nucleation temperature $T_n \sim 0.98 T_c$ is much smaller than the upper limit and the transition rate $\frac{\beta}{H} = 17742$ is large typically due to small supercooling. In addition, $\alpha = 0.298$ gives relativistic bubbles $v_w = 0.86$ and corresponding parameters are listed in Table II. Figure 1 shows the GW spectrum in the pure gluon system, whose peak frequency is about $10^{-3}$ Hz and peak energy density is about $10^{-11}$. It is only detectable for the original LISA and far away from the nanohertz band. The sensitivity curves of GWs detectors are from Ref. [85] (the same below).

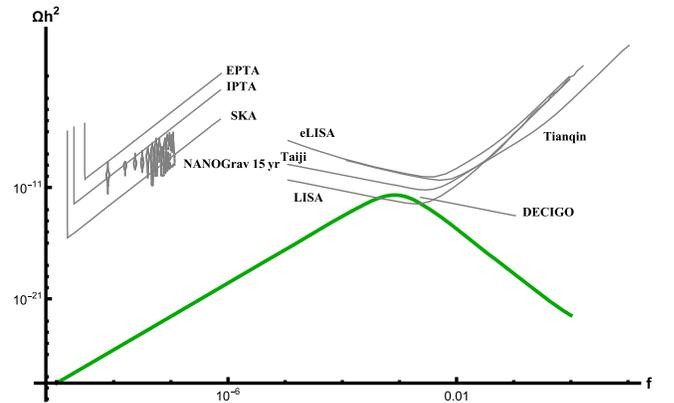

FIG. 1. The GW spectrum in the pure gluon system.

## IV. THE DECONFINEMENT PHASE TRANSITION IN THE FRIEDBERG-LEE MODEL

The Lagrangian of the Friedberg-Lee (FL) model has the form of [12–14]

$$\mathcal{L}_{\mathrm{FL}} = \bar{\Psi}(i\slashed{\partial} - g\phi)\Psi + \frac{1}{2}\partial_\mu\phi\partial^\mu\phi - U_{\mathrm{FL}}(\phi) \quad (27)$$

with $\phi$ the scalar field and $\Psi$ the quark field. The potential $U_{\mathrm{FL}}(\phi)$ takes the Ginzburg-Landau type

$$U_{\mathrm{FL}}(\phi) = \frac{1}{2!}a\phi^2 + \frac{1}{3!}b\phi^3 + \frac{1}{4!}c\phi^4. \quad (28)$$

Following Ref. [86], here we choose $a = 0.68921$ GeV$^2$, $b = -287.59$ GeV, $c = 20000$ and $g = 12.16$, which can successfully reproduce the static properties of nucleon. Similar to the MIT model, there exists a soliton solution at $T < T_c$ which serves as a "bag" to confine quarks, while only a damping oscillation solution exists at $T > T_c$, in which quarks are set free. Thus the FL model can be used to describe a deconfinement phase transition.

Considering the one-loop finite temperature contribution, the effective grand potential is [87,88]

$$\begin{aligned}\Omega_{\mathrm{FL}}(\phi, T, \mu) = U_{\mathrm{FL}}(\phi) + T \int \frac{d^3\vec{p}}{(2\pi)^3} &\{\ln(1 - e^{-E_\phi/T}) \\ - \nu[\ln(1 + e^{-(E_\Psi - \mu)/T}) \\ + \ln(1 + e^{-(E_\Psi + \mu)/T})]\}\end{aligned} \quad (29)$$

with $\nu = 2N_f N_c = 2 \times 2 \times 3 = 12$.

The effective mass of quark and the scalar field $\phi$ are $m_\Psi = g\phi$ and $m_\phi^2 = a + b\phi + \frac{c}{2}\phi^2$ respectively, then the energy is $E_i = \sqrt{\vec{p}^2 + m_i^2}$ ($i = \phi, \Psi$). The order parameter $\phi$ can be determined by solving the gap equation

$$\frac{\partial\Omega_{\mathrm{FL}}}{\partial\phi} = 0. \quad (30)$$

Figure 2(a) shows the effective grand potential of the FL model as the chemical potential $\mu$ increases from 0 to 300 MeV. Potential curves except for $\mu = 300$ MeV are plotted at corresponding $T_c$, while the curve for $\mu = 300$ MeV is plotted at $T = 0.01$ MeV and does not have two degenerate vacua, suggesting that the phase transition cannot happen at a finite temperature if $\mu$ is too large. The potential barrier exists regardless of $\mu$ and rises with increasing $\mu$. The order parameter $\phi$ jumps to around 0.037 GeV from 0 at $T_c$, and all solutions to the gap equation indicate a first-order phase transition as shown in Fig. 2(b).

The chemical potential $\mu$ barely changes the expected value of $\phi$, but $T_c$ smoothly drops from 0.12 GeV to around 0.03 GeV for $0 < \mu < 280$ MeV and sharply falls to 0 at the critical chemical potential $\mu_c = 297.5$ MeV, as shown in Fig. 3(a). The phase transition is always of first order without a CEP when $\mu < 297.5$ MeV, but cannot start when $\mu > 297.5$ MeV because the two vacua can never degenerate as displayed in Fig. 2(a). Besides, when $\mu$ is in the left adjacent region of 297.5 MeV, although the two vacua can degenerate, the phase transition cannot complete because the nucleation possibility is too small. Thus, there must exist a point where the nucleation criterion is barely reached, as the red triangle indicates in Fig. 3(a) at $\mu = 287.55$ MeV. This point is called the critical nucleation point (CNP) for convenience in the following texts.

Data of $T_c$ are listed in Table III, which can be numerically fitted by an even power function when the chemical potential $\mu$ is not so large, e.g., $\mu < 287.5$ MeV. More and more higher-order terms must be included as $\mu$ approaches 297.5 MeV from the CNP. The function form

$$\begin{aligned}T_c/\mathrm{GeV} = 0.11974 &- 0.93782(\mu/\mathrm{GeV})^2 \\ &- 2.1671(\mu/\mathrm{GeV})^4\end{aligned} \quad (31)$$

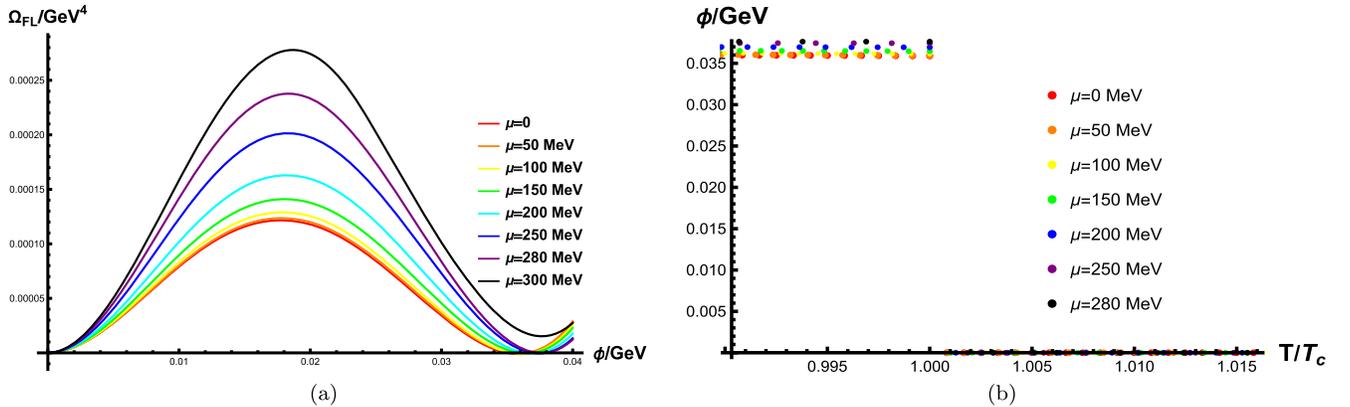

FIG. 2. Panel (a) shows the effective grand potential with different chemical potential $\mu$ at $T_c$ in the FL model. Panel (b) shows the order parameter $\phi$ as a function of $T/T_c$ with different chemical potential $\mu$ in the FL model.

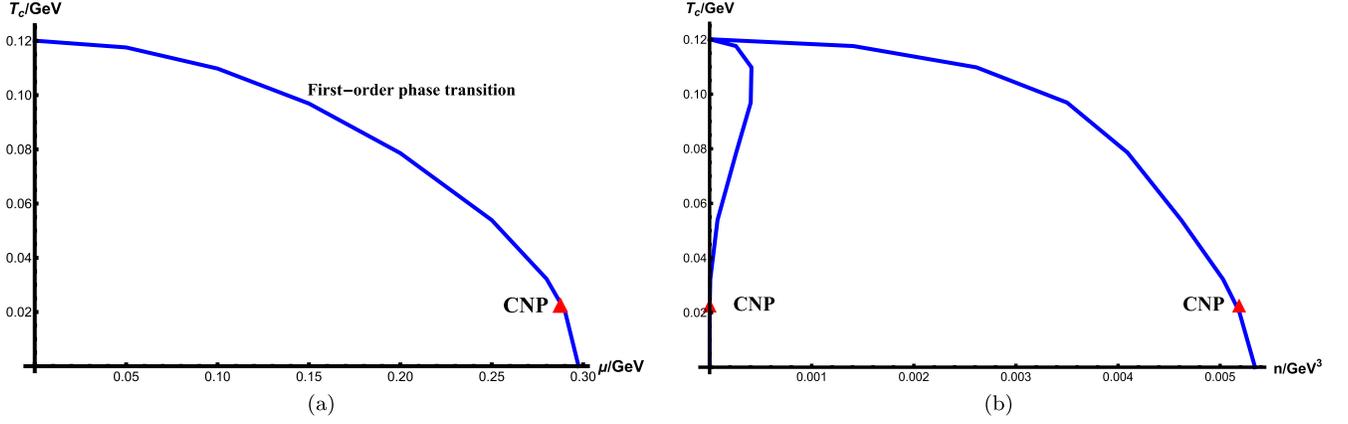

FIG. 3. Panel (a) shows the phase diagram in the FL model in terms of temperature $T$ and chemical potential $\mu$. No CEP exists. Panel (b) shows the phase diagram in the FL model in terms of temperature $T$ and number density $n$.

is similar to the behavior of $T_c$ in the PNJL model under rotation [89] when $\mu$ is not so large, which implies that the impact on QCD systems from the chemical potential $\mu$ may be similar to the angular velocity.

We select three cases with different chemical potentials $\mu$ to exemplify the bounce action in the FL model in Fig. 4. In each case, the bounce action reaches 180 at small supercooling of about 1%, and in the weak supercooling case, the bounce action curves can be uniformly fitted by

$$\frac{S_3}{T} = \frac{a}{\eta^2} + c, \quad (32)$$

where $\eta(T) = 1 - \frac{T}{T_c}$ measures the supercooling, and parameters $a$ and $c$ are dictated by the effective grand potential. $\eta(T_n)$ can be estimated by

$$\eta = \sqrt{\frac{a}{180 - c}} \quad (33)$$

and thus, the parameter $\beta/H$ is

$$\frac{\beta}{H} = 2a\left(\frac{1}{\eta^3} - \frac{1}{\eta^2}\right) = 2(180 - c)\left(\frac{1}{\eta} - 1\right). \quad (34)$$

Bounce action data (blue dots) and the fitting function (red, orange and magenta lines) fit well in Fig. 4. The trend in Fig. 4 illustrates that a larger chemical potential $\mu$ as well as a consequent higher potential barrier blunts the nucleation process, and phase transitions need stronger supercooling to start. From the perspective of the bounce action, curves with larger $\mu$ are higher than others at the same supercooling and need stronger supercooling to reach 180. Therefore, phase transitions are slower with larger $\mu$, i.e., values of $\beta/H$ are smaller. According to Eq. (34), $\beta/H$ is approximately $10^4$ and proportional to $\eta^{-1}$ if $c \ll 180$ with weak supercooling $\eta \sim 1\% \ll 1$ as shown in Fig. 4.

Equation (32) is similar to the thin wall approximation form [67] except for the constant $c$ which is mainly dictated by the existence of the potential barrier. For example, Fig. 4 in Ref. [90] and Fig. 1 in Ref. [91] show the duration of the potential barrier in the PQM and QM model, the potential barrier exists longer or becomes permanent as $\mu$ moves from the CEP to the CNP while $c$ keeps increasing from $c \ll 180$ to $c \sim 100$. If the potential barrier vanishes at small $\eta$, the curve drops directly from infinity to $S_3/T \ll 180$ quickly and in this situation, which usually can be observed near the CEP, typically slight supercooling $\eta$ is given by $a \ll 1$ and $c < 0$, the bounce action can always reach 180 and $\beta/H \sim 10^{4-5}$ is very large. If the potential barrier does not vanish or vanishes at very low temperature, the shape of the curve is a "U" type [92], the bounce action drops from infinity at $\eta = 0$ to a positive minimum around $\eta \sim 0.5$ at which the thin wall approximation expires, then increases as supercooling becomes further strong due to "thick" wall effect. If the minimum is much smaller than 180, which can be observed away from both the CEP and the CNP in the phase diagram and occupies most of the region that supports a first-order phase transition. Typically weak supercooling $\eta$ as well as large $\beta/H \sim 10^3$ is given by $a \sim 1$ and $0 < c \ll 180$. Furthermore, in the region closer to the CNP, the minimum of the bounce action is in the same order of 180, supercooling $\eta$ approaches 0.4 with $a \gg 1$ and $c \sim 100$, $\beta/H$ is around $10^2$. To obtain small transition rate $\beta/H \sim 10$, approximately $c$ should be slightly smaller than 180 and corresponding $\eta$ is around 0.5. Actually Eq. (32) works well at $\eta < 0.4$ or $\beta/H > 10^3$ but deviates from the bounce action when $\eta \sim 0.5$ or so. To establish a quick

TABLE III. The critical temperature $T_c$ with different chemical potential $\mu$ in the FL model.

| $\mu$/GeV | 0 | 0.05 | 0.1 | 0.15 | 0.2 | 0.25 | 0.28 |
|---|---|---|---|---|---|---|---|
| $T_c$/MeV | 120.1 | 117.6 | 109.8 | 96.9 | 78.6 | 53.9 | 32.2 |

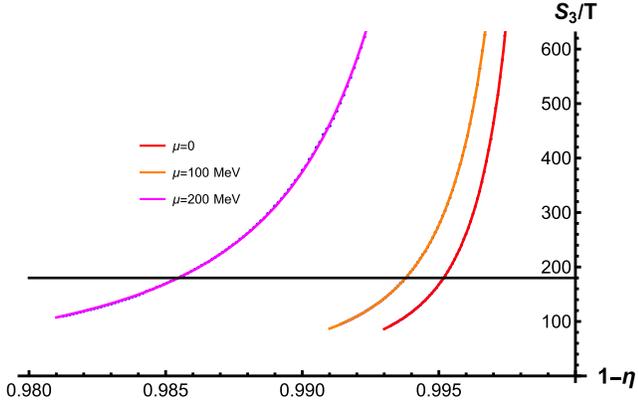

FIG. 4. The bounce action (blue dots) and the fitting function (red, orange and magenta lines) with different chemical potential $\mu$ in the FL model. The black line is for 180.

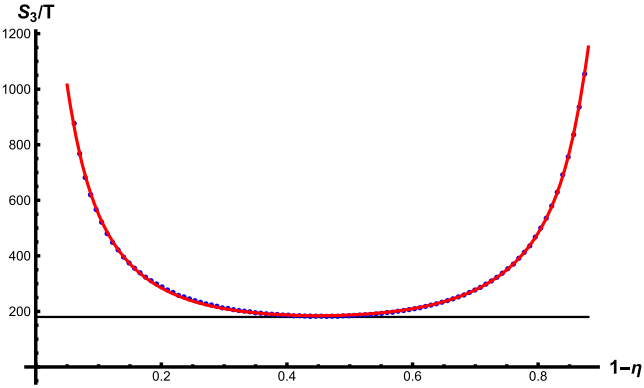

FIG. 5. The bounce action (blue dots) and the numerical function (red) in the CNP case as an example. The black line is for 180.

estimate of nucleation near the CNP, the full bounce action curve can be fitted by

$$\frac{S_3}{T} = \frac{a}{\eta^2} + \frac{b}{(1-\eta-d)^2} + c. \quad (35)$$

The bounce action at the CNP is plotted in Fig. 5 as an example, the blue dots and the red line fit well. In this CNP case, the minimum of $S_3/T$ is 182, barely touching 180. Numerically $d$ is small and hence the minimum is reached at around $\eta \sim 0.5$. Accurate values of $a$ and $c$ are also decided by specific models. In the situation to the right of the CNP, the minimum is larger than 180, and the thermal decay statistically does not complete. Vacuum nucleation at extreme supercooling where $S4$ should be adopted instead of $S_3/T$ is not taken into consideration because it is too cryogenic for common QCD phase transitions.

Using Eqs. (32) and (34), parameters $T_n$ and $\beta/H$ can be calculated conveniently. Table IV lists parameters $T_n$, $\alpha$ and $\beta/H$ in the FL model to calculate GWs and $\beta/H$ (blue dots) as well as its fitting function plotted in Fig. 6. Since no CEP exists in the FL model, values of $\beta/H$ decrease from a finite value around $10^5$ to $10^3$ when $\mu$ increases from 0 to 280 MeV and sharply falls from $10^3$ to 0 in a very narrow interval near CNP indicated by the red triangle in Fig. 6. To the right of the CNP, $\beta/H$ for thermal decay keeps zero until $\mu_c$ as indicated by the red square in Fig. 6, thus the phase transition cannot complete until vacuum decay dominates at extreme low temperature, which is beyond our consideration here. Considering the constraint $\beta/H \sim 10$ on nanohertz GWs, only a 0.1 MeV wide interval of $\mu$ to the left of the CNP is eligible. Behavior of $\beta/H$ resembles $T_c$ and can also be fitted in terms of $\mu$ scaled by $\mu_c$

$$\ln\left(\frac{\beta}{H}\right) = -a\ln\left(\frac{\mu - \mu_{\text{CEP}}}{\mu_c}\right) + b\ln\left(\frac{\mu_{\text{CNP}} - \mu}{\mu_c}\right) + c, \quad (36)$$

where $a = 20.1$, $b = 1.04$ and $c = -74.5$ are model dependent parameters in the FL model. For models without the CEP like the FL model, $\mu_{\text{CEP}}$ is substituted with an effective negative value, e.g., $-21.2$ GeV in the FL model. In addition, values of $\alpha$ are all around 0.2 due to dense baryons $\mu n \gg p$ and slightly increase with increasing $\mu$.

Figure 7 plots the GW spectra in the FL model. When the chemical potential $\mu$ is small, e.g., $\mu \leq 200$ MeV, the peak frequency is around $10^{-3}$ Hz and the peak energy density is around $10^{-12}$. As $\mu$ increases, the peak frequency declines while the peak energy density rises. When $\mu$ is as large as 280 MeV, the peak frequency becomes $10^{-6}$–$10^{-5}$ Hz and the peak energy density rises to around $10^{-10}$–$10^{-9}$. GWs from most of the first-order phase transition region are in the LISA and Taiji band but beyond the ranges of current detectors, except in a very narrow interval close to the CNP, 287.5 MeV > $\mu$ > 287.55 MeV, where $\beta/H \sim 10^1$. Only in this narrow interval does the peak frequency decrease to as low as $10^{-9}$–$10^{-7}$ Hz and the peak energy density rise to the detectable interval $10^{-9}$–$10^{-8}$, coinciding with nanohertz GW data.

## V. THE CHIRAL PHASE TRANSITION IN THE QM AND PQM MODEL

The effective Lagrangian of the two-flavor quark-meson (QM) model is [91,93]

$$\mathcal{L}_{\text{QM}} = \frac{1}{2}\partial^\mu\sigma\partial_\mu\sigma + \frac{1}{2}\partial^\mu\vec{\pi}\partial_\mu\vec{\pi} + \bar{\Psi}i\partial\!\!\!/\Psi - g\bar{\Psi}(\sigma + i\gamma_5\vec{\tau}\cdot\vec{\pi})\Psi - U_{\text{QM}}(\sigma,\vec{\pi}), \quad (37)$$

in which the Yukawa coupling term gives the interaction between quarks $\Psi = (u,d)$ and scalar mesons, including three pions $\vec{\pi}$ and one $\sigma$ meson, and $\vec{\tau}$ are the Pauli matrices. The last term in the above equation is the pure mesonic potential given by the expression

TABLE IV. Parameters $T_n$, $\beta/H$, and $\alpha$ with different chemical potential $\mu$ in the FL model.

| $\mu$/GeV | 0 | 0.05 | 0.1 | 0.15 | 0.2 | 0.25 | 0.28 | 0.287 | 0.2875 |
|---|---|---|---|---|---|---|---|---|---|
| $T_n$/MeV | 119.63 | 117.07 | 109.26 | 96.13 | 77.63 | 52.20 | 28.38 | 19.70 | 11.70 |
| $\beta/H$ | 74948 | 69108 | 57255 | 41888 | 24178 | 9412 | 1966 | 242 | 44 |
| $\alpha$ | 0.184 | 0.187 | 0.196 | 0.210 | 0.230 | 0.251 | 0.259 | 0.260 | 0.269 |

$$U_{\mathrm{QM}}(\sigma, \vec{\pi}) = \frac{\lambda}{4}(\sigma^2 + \vec{\pi}^2 - v^2)^2 - H\sigma, \quad (38)$$

The chiral symmetry spontaneously breaks when $\sigma$ takes a nonzero vacuum expectation value, which is equal to the decay constant of pions, $\sigma = f_\pi = 93$ MeV. This results in three Goldstone pions which are massless in the chiral limit, but they become massive when the chiral symmetry is explicitly broken. Meanwhile, the quarks get constituent masses $m_\Psi = gf_\pi$ with $g = 3.3$ if assuming that $m_\Psi$ contributes to one-third of the nucleon mass. Partially conservation of axial current gives the parameter $H = f_\pi m_\pi^2$, where $m_\pi = 138$ MeV is the pion mass due to the nonzero current quark mass. Another parameter $\lambda$ is determined by the expression for the mass of the $\sigma$ meson, $m_\sigma^2 = m_\pi^2 + 2\lambda f_\pi^2 = (500 \text{ MeV})^2$, and then $v^2 = f_\pi^2 - \frac{m_\pi^2}{\lambda}$ is also fixed.

The QM model can be promoted to simultaneously describe both the chiral and the deconfinement phase transitions with the inclusion of the Polyakov loop variables. In such an extension, the quarks are coupled with gauge fields through the covariant derivative $\partial_\mu \longrightarrow D_\mu = \partial_\mu - i\delta_{0\mu}A_0$ [94] and the effective Lagrangian of two-flavor Polyakov quark-meson (PQM) model is

$$\mathcal{L}_{\mathrm{PQM}} = \mathcal{L}_{\mathrm{QM}} + \bar{\Psi}\gamma_0 A_0 \Psi + \mathcal{U}_{\mathrm{PQM}}(\Phi, \bar{\Phi}, T), \quad (39)$$

where $\mathcal{U}_{\mathrm{PQM}}(\Phi, \bar{\Phi}, T)$ is the effective potential of the Polyakov loop. The fitting form in Sec. III does not work accurately here due to the temperature being much lower than $T_0$, hence a more complicated fitting form [82] is adopted

$$\frac{\mathcal{U}_{\mathrm{PQM}}(\Phi, \bar{\Phi}, T)}{T^4} = -\frac{b_2(T)}{2}\Phi\bar{\Phi} - \frac{b_3}{6}(\Phi^3 + \bar{\Phi}^3) + \frac{b_4}{4}(\Phi\bar{\Phi})^2 \quad (40)$$

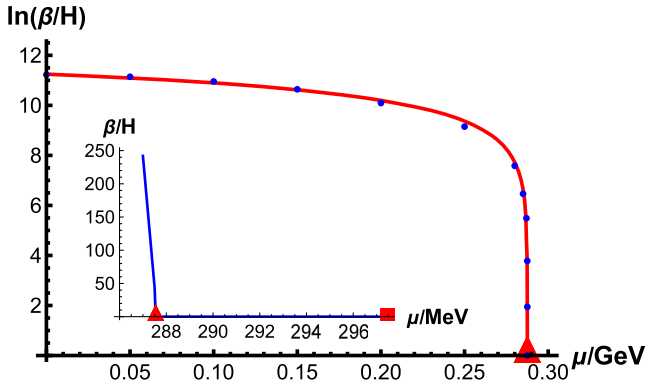

FIG. 6. $\beta/H$ (blue dots) and the fitting function (red) with different chemical potential $\mu$ in the Friedberg-Lee model.

with a temperature dependent coefficient

$$b_2(T) = a_0 + a_1\left(\frac{T_0}{T}\right) + a_2\left(\frac{T_0}{T}\right)^2 + a_3\left(\frac{T_0}{T}\right)^3, \quad (41)$$

whose parameters are listed as follows:

$$a_0 = 6.75, \quad a_1 = -1.95, \quad a_2 = 2.625,$$
$$a_3 = -7.44, \quad b_3 = 0.75, \quad b_4 = 7.5. \quad (42)$$

Considering the back action of the dynamical quarks on the gluonic sector, in the two-flavor case $T_0 = 208$ MeV is used instead of $T_0 = 276$ MeV for pure $SU(3)$ gauge theory.

In the mean field approximation, the effective grand potential of the PQM model is

$$\Omega_{\text{PQM}} = U_{\text{QM}}(\sigma, \vec{\pi}) + \mathcal{U}_{\text{PQM}}(\Phi, \bar{\Phi}, T) - \nu \int \frac{d^3\vec{p}}{(2\pi)^3} \bigg\{ E_\Psi$$

$$+ \frac{T}{N_c} \bigg[ \ln\bigg(1 + 3\bigg(\Phi e^{-\frac{E_\Psi - \mu}{T}} + \bar{\Phi} e^{-2\frac{E_\Psi - \mu}{T}}\bigg) + e^{-3\frac{E_\Psi - \mu}{T}}\bigg)$$

$$+ \ln\bigg(1 + 3\bigg(\bar{\Phi} e^{-\frac{E_\Psi + \mu}{T}} + \Phi e^{-2\frac{E_\Psi + \mu}{T}}\bigg) + e^{-3\frac{E_\Psi + \mu}{T}}\bigg)\bigg]\bigg\}.$$

(43)

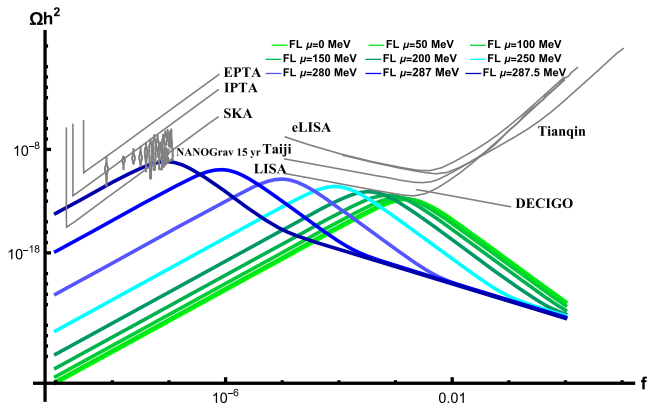

FIG. 7. GW spectra with different chemical potential $\mu$ in the FL model.

The third term is the vacuum one-loop contribution and diverges severely, it can be written in a finite form after renormalization [95]

$$-\frac{N_f N_c}{8\pi^2} g^4 \sigma^4 \ln\bigg(\frac{\sigma}{f_\pi}\bigg). \tag{44}$$

By taking the limits as $\Phi = \bar{\Phi} \to 1$, the effective grand potential backs the QM model

$$\Omega_{\text{QM}} = U_{\text{QM}}(\sigma, \vec{\pi}) - \nu \bigg\{ \int \frac{d^3\vec{p}}{(2\pi)^3} E_\Psi$$

$$+ T \int \frac{d^3\vec{p}}{(2\pi)^3} \bigg[\ln\bigg(1 + e^{-\frac{E_\Psi - \mu}{T}}\bigg) + \ln\bigg(1 + e^{-\frac{E_\Psi + \mu}{T}}\bigg)\bigg]\bigg\}.$$

(45)

Order parameters in the PQM model $\sigma$, $\Phi$ and $\bar{\Phi}$ can be calculated by solving the gap equations

$$\frac{\partial\Omega}{\partial\sigma} = \frac{\partial\Omega}{\partial\Phi} = \frac{\partial\Omega}{\partial\bar{\Phi}} = 0, \tag{46}$$

and it reduces to the QM model if setting $\Phi = \bar{\Phi} = 1$ and solving $\frac{\partial\Omega}{\partial\sigma} = 0$ only.

The potential curves in the QM model are plotted at the corresponding $T_c$ in Fig. 8(a). Increasing chemical potential $\mu$ not only raises the potential barrier but also stretches it, both retarding the phase transition. In the PQM model the potential has the same properties as in the QM model, but three order parameters make the four-dimensional tunneling path $\Omega(\sigma, \Phi, \bar{\Phi})$ too complicated to calculate the bounce action. To simplify, $\sigma$ is chosen as the main parameter and for each value of $\sigma$ between the true vacuum and the false vacuum, solving $\frac{\partial\Omega}{\partial\Phi} = \frac{\partial\Omega}{\partial\bar{\Phi}} = 0$ gives two continuous functions $\Phi(\sigma)$ and $\bar{\Phi}(\sigma)$, thus the tunneling path can be numerically reconstructed with respect to one parameter only $\Omega(\sigma, \Phi(\sigma), \bar{\Phi}(\sigma))$.

Figure 8(b) shows an example of the potential in the PQM model, where all curves are plotted as functions of $\sigma$ so that the four-dimensional potential is projected on a two-dimensional form. Values of $\Phi$ and $\bar{\Phi}$ of the blue curve and the red curve are different and determined so that the minimum attained by the blue curve is the true vacuum and the lowest false vacuum by the red curve, respectively. The gray curve connecting the two vacua is the tunneling path reconstructed.

The phase diagrams of the QM model are plotted in Figs. 9(a) and 9(b), where the CEP at $\mu = 299.4$ MeV and $T_c = 0.03217$ GeV is indicated by the red star and the CNP at $\mu = 309.6$ MeV by the red triangle. Transition in the left region of CEP is a crossover and turns to a first-order phase transition after the advent of potential barrier at the CEP with increasing $\mu$. Compared with the FL model, only a 10 MeV tight interval of $\mu$ supports a first-order phase transition and it can be predicted that the relative parameters are more sensitive to $\mu$. The critical temperature $T_c$ of the first-order phase transition drops almost linearly from over 0.03 to 0.01 GeV in a 10 MeV interval from $\mu = 299.4$ MeV to $\mu = 309.6$ MeV and rapidly approaches 0 at $\mu_c = 311.1$ MeV.

The bounce action in the PQM and QM model is shown in Fig. 10 and numerically fitted by the same functional form as in the FL model. Comparing Figs. 10(a) and 10(b), we can find that the bounce action in both models reaches 180 at small supercooling, but supercooling is smaller and $\beta/H$ is larger in the QM model than in the PQM model, which means the presence of gluons facilitates a decrease of the bounce action and makes the phase transition easier.

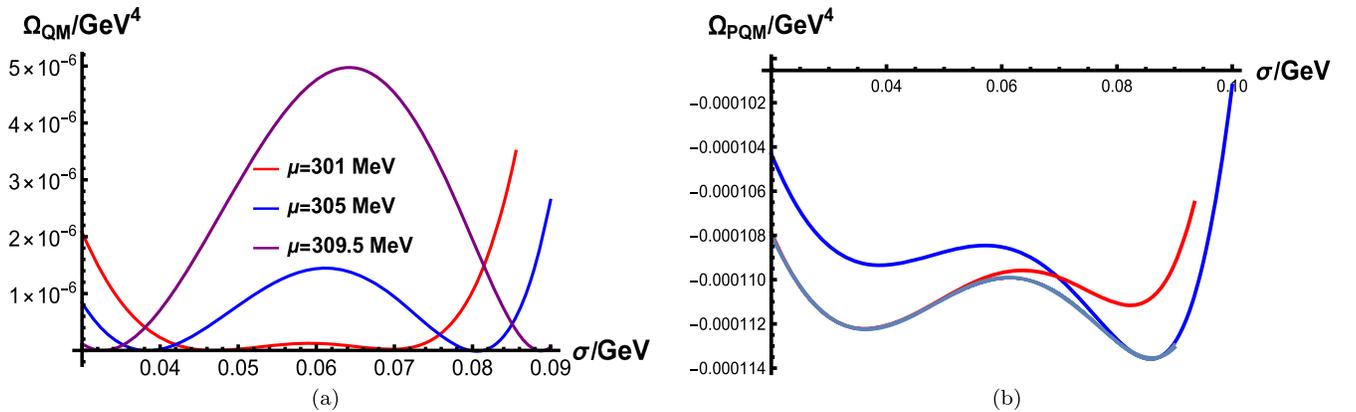

FIG. 8. Panel (a) shows the effective grand potential with different chemical potential $\mu$ at $T_c$ in the QM model. Panel (b) shows the effective grand with $\mu = 306$ MeV at $T_n$ in the PQM model.

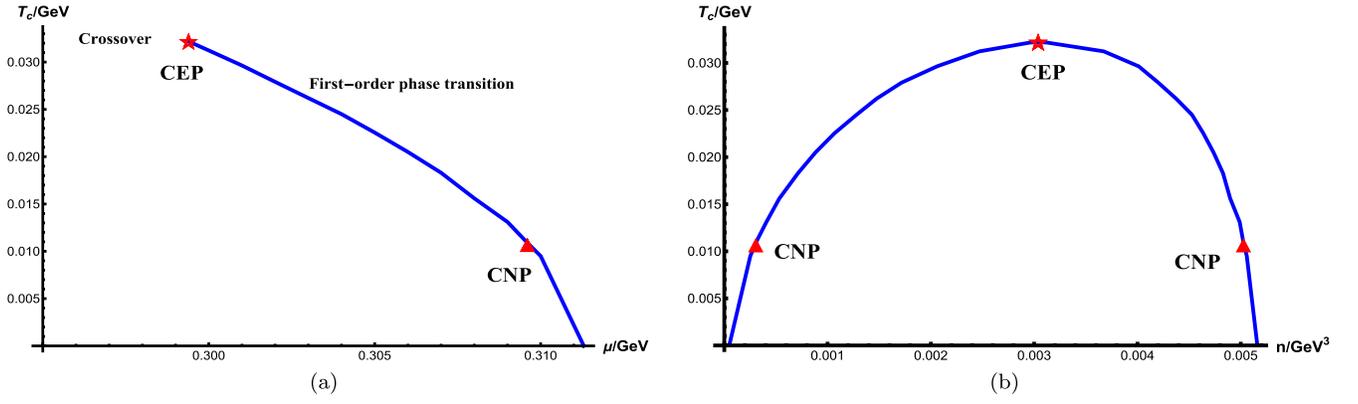

FIG. 9. Panel (a) shows the phase diagram in the QM model in terms of $T$ and $mu$. The red star is for CEP at $\mu = 299.4$ MeV and triangle for $\mu = 309.6$ MeV. $T_c$ approaches 0 at $\mu = 311.3$ MeV. Panel (b) shows the phase diagram in the QM model in terms of $T$ and $n$. The red star is for CEP at $\mu = 299.4$ MeV and triangle for $\mu = 309.6$ MeV.

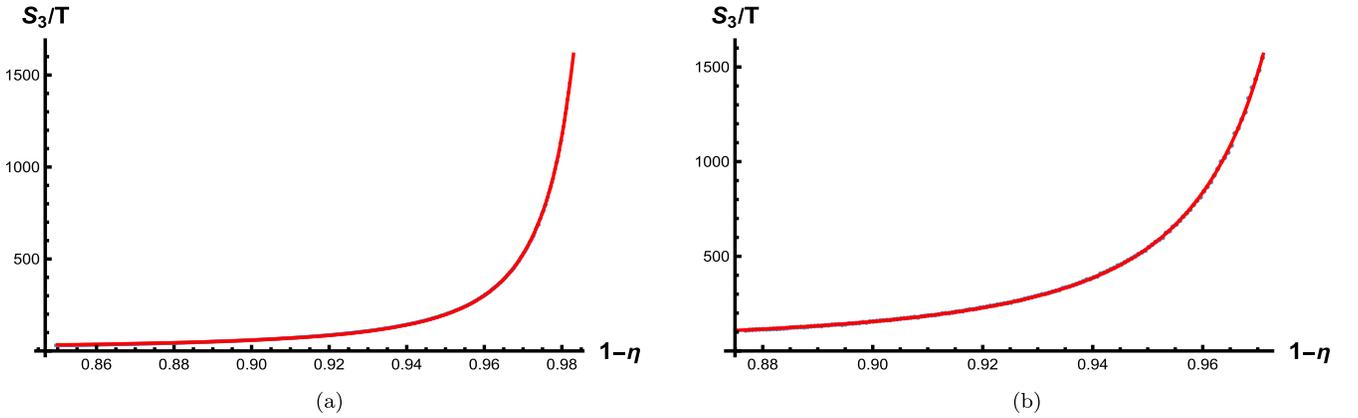

FIG. 10. The bounce action (blue dots) and its numerical functions (red) with $\mu = 308$ MeV in the PQM (panel a) and QM (panel b) model.

Except for the slight difference between the QM and PQM model mentioned above, they share similar behaviors and thus our discussion mainly focuses on the QM model. We select $\mu = 306, 307, 308, 309, 309.5, 309.59$ MeV in the QM model and $\mu = 306, 308$ MeV in the PQM model as typical examples. Parameters for GW calculation are listed in Table V and $\beta/H$ (blue dots) as well as its fitting function (red line) are plotted in Fig. 11. Different from the FL model in which $\beta/H$ starts to decrease from a finite value 74948 from $\mu = 0$, $\beta/H$ in the QM model starts to fall from infinity from the CEP $\mu = 299.4$ MeV indicated by the red star and soon reaches a plateau during which $\beta/H$ varies from $10^4$ to $10^3$ and sharply falls from $10^3$ to 0 in a very narrow interval near the CNP indicated by the red triangle. To the right of the CNP, $\beta/H$ keeps zero until $\mu_c$ indicated by the red square. Considering the constraint $\beta/H \sim 10$, again only a 0.1 MeV wide interval to the left of CNP is eligible. $\beta/H$ can be fitted by the same form as Eq. (36) with $a = 1.59$, $b = 0.980$ and $c = 9.64$. Moreover, parameter $\alpha$ is also around 0.2 and increases with $\mu$.

TABLE V. Parameters $T_n$, $\beta/H$, and $\alpha$ with different chemical potential $\mu$ in the PQM and QM model.

| | PQM | | QM | | | | | |
|---|---|---|---|---|---|---|---|---|
| $\mu$/MeV | 306 | 308 | 306 | 307 | 308 | 309 | 309.5 | 309.59 |
| $T_n$/MeV | 45.26 | 35.72 | 19.93 | 17.40 | 14.52 | 10.69 | 7.20 | 6.20 |
| $\beta/H$ | 17574 | 6341 | 11888 | 6130 | 3179 | 1095 | 208 | 24 |
| $\alpha$ | 0.217 | 0.234 | 0.206 | 0.218 | 0.229 | 0.238 | 0.245 | 0.247 |

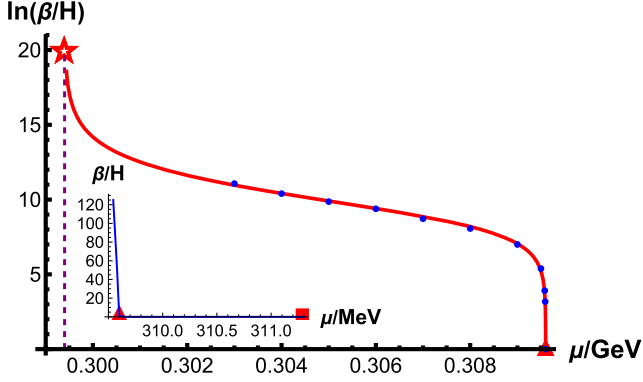

FIG. 11. $\beta/H$ (blue dots) and the fitted function (red) with different chemical potential $\mu$ in the QM model.

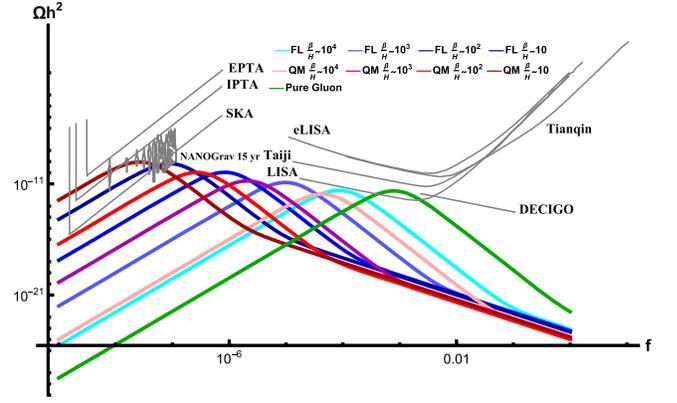

FIG. 13. GW spectra in different models.

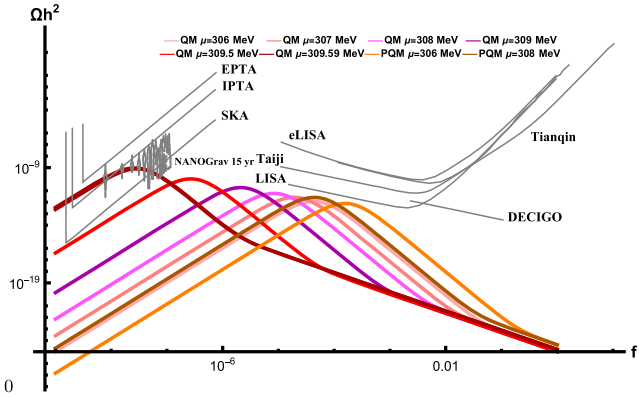

FIG. 12. GW spectra with different chemical potential $\mu$ in the QM and PQM model.

Figure 12 shows GW spectra in the PQM and QM model. When $\mu$ is small, e.g., $\mu < 309.5$ MeV, GW spectra lie beyond the detection range of current detectors and mostly concentrate around $10^{-6}$ Hz to $10^{-4}$ Hz, the Taiji and LISA band. As $\mu$ increases, peak frequency decreases and peak energy density increases. Similar to the situation in the FL model, only in a narrow interval 309.6 MeV $> \mu >$ 309.5 MeV where $\beta/H \sim 10^1$, the peak frequency is around $10^{-9}$ and is detectable for SKA, IPTA or EPTA, coinciding with nanohertz GW data.

## VI. DISCUSSION AND SUMMARY

In this work, we calculate the GW spectra and the corresponding phase transition strength and transition rate from first-order phase transitions in several QCD models. It is noticed that in most cases, the transition rate is large, and $\beta/H \gg 10$, except for those near the CNPs where one can obtain $\beta/H \sim 10$. In Fig. 13, we summarize the GW spectra from QCD phase transitions with transition rates on the order of $\beta/H \sim 10^1, 10^2, 10^3, 10^4$ in the FL model and QM model at different finite chemical potentials, with detailed parameters shown in Table VI. The GWs from the pure gluon system at high temperature and zero chemical potential with $\beta/H \sim 10^4$ are used as a reference, and they lie in the range of the LISA detector window. As the chemical potential $\mu$ increases in different models, $\beta/H$ becomes smaller, and the peak frequency of the GW spectra gradually moves from the LISA band to the nanohertz band. However, most of the spectra are beyond the detection of Taiji or other currently planned detectors except for those with $\beta/H \sim 10$, which coincide with nanohertz GW data. Therefore, it can be expected that most signals from QCD first-order phase transitions fall within the range of Taiji but are unlikely to be located in the nanohertz band. In the PQM and QM models, signals are suppressed by small $\alpha$ due to dense baryons; however, a sparser background medium stirred by QCD first-order phase transitions is able to produce GWs that fall right within the range of more sensitive detectors such as the original LISA.

As mentioned before, a larger chemical potential $\mu$ results in a larger bounce action for the same supercooling and a smaller transition rate in the same model. In other words, a larger $\mu$ leads to slower phase transitions and a lower peak frequency of the GW spectra. To some extent, the influence of $\mu$ can be understood as follows. As emphasized in Ref. [95], it is necessary to include a one-loop Coleman-Weinberg-type logarithmic correction term, such as in Eq. (44), as a counterterm to cancel the Coleman-Weinberg-type logarithmic high-temperature thermal correction term at zero chemical potential $\mu = 0$. However, this cancellation becomes incomplete if the finite density correction is included, which facilitates the appearance of a potential barrier, as seen in the QM and PQM models. Mathematically, the finite density correction term depends on the dispersion relation. For models with a dispersion relation like $E = \sqrt{p^2 + g^2\sigma^2}$, the finite density correction introduces high-order power terms and Coleman-Weinberg-type logarithmic terms, which the tree-level potential does not contain. These correction terms favor the appearance of a barrier in a crossover or enhance the barrier in a first-order phase transition.

TABLE VI. Parameters for Fig. 13.

|  | FL | | | | QM | | | | Pure gluon |
|---|---|---|---|---|---|---|---|---|---|
| $\mu$/MeV | 250 | 280 | 287 | 287.5 | 306 | 309 | 309.5 | 309.59 | 0 |
| $\beta/H$ | 9412 | 1966 | 242 | 44 | 11888 | 1095 | 208 | 24 | 17742 |
| $\alpha$ | 0.251 | 0.259 | 0.260 | 0.269 | 0.206 | 0.238 | 0.245 | 0.247 | 0.298 |

We also offer an intuitive understanding on this result. From the experience in the NJL model, it is learned that if there is a repulsive vector interaction in the Lagrangian, the system will develop an effective chemical potential $\mu_*$, e.g., see Refs. [96–98]. This stimulates us to understand the effect of $\mu$ as an effective repulsive vector interaction. For a system at high baryon chemical potential, the attractive interaction in the scalar channel induces a dynamical quark mass and develops a potential well, and the baryon chemical potential or the effective repulsive interaction induces a potential barrier. Similar to the nuclear force including the repulsive interaction, when the chemical potential or the effective repulsive interaction increases, the potential barrier can be formed, and a first-order phase transition will start to occur. When the chemical potential or the effective repulsive interaction further increases, the height of the potential barrier enhances, thus giving a stronger first-order phase transition.

The influence of the finite density term on the potential can be directly reflected in the bounce action and $\beta/H$. The process can be tracked as follows as the chemical potential gradually increases. The transition is an ephemeral crossover if $\mu$ is zero or small according to lattice data because the potential has no barrier; hence, the supercooling is 0, and $\beta/H$ can be treated as infinity. However, increasing the chemical potential can delay the transition by raising the potential. Eventually, the emergence of a potential barrier at the CEP smoothly changes a crossover into a first-order phase transition, passing through a second-order phase transition at the CEP. Due to the smooth transition, near the CEP the potential barrier is not high enough, and the phase transition completes in an extremely short time; supercooling is slightly larger than 0, while $\beta/H$ is huge but no longer infinite. For models that have a potential barrier regardless of the magnitude of the chemical potential, e.g., the FL model, $\beta/H$ starts decreasing from a finite value at $\mu = 0$. If the chemical potential further increases, the potential barrier rises and slows down the phase transition; stronger supercooling is needed for the bounce action to reach 180, and $\beta/H$ gradually decreases to $10^3$. Naively, if

$\mu$ continues to increase, $\beta/H$ becomes as small as 10 at a specific $\mu$ near the CNP, but in that region slight changes in $\mu$ can trigger a sharp drop in $T_c$, and the nucleation temperature soon becomes lower than the nucleosynthesis temperature. Besides, the error for that value of $\mu$ must not exceed about one ten-thousandth; e.g., 309.6 MeV $> \mu >$ 309.58 MeV, which is unreasonable in the early universe considering stochastic fluctuations.

We propose the critical nucleation point (CNP) to describe the critical point at which the nucleation criterion is barely reached. In this work, nucleation is described by the nucleation temperature $T_n$, at which $\Gamma/H^4 \sim 1$. This criterion itself is flexible and not unique, e.g., the percolation temperature $T_p$ is an alternative, thus the real location of the CNPs has slight uncertainty. But near the CNPs, the phase diagram is sensitive to $\mu$, but not to $T$; hence, the uncertainty of the nucleation condition does not change the existence of the CNPs or other physical facts. To the right of the CNPs, the nucleation possibility is suppressed, and the bounce action cannot reach the nucleation criterion. Thermal decay cannot complete, and the PQNs would remain in the early universe. Near the CNPs in the FL and QM models, we have $\mu/T \sim 10$–100, which will be proper after a little inflation. These dense quark nuggets, inlaid in the later universe, serve as seeds of primordial compact stars or future cosmic structures. In addition, resembling the electroweak phase transitions with an extreme supercooling mechanism, when the universe cools down to a strongly supercooled state for the QCD phase transition, e.g., much lower than BBN, vacuum decay prevails, and the phase transition can still slowly complete much later, turning dense quark matter into normal baryonic matter. The phase transition would erupt most parts of the baryonic matter, with cold, sparse matter remaining. Because of the slow phase transition, bulky true vacuum bubble collisions may generate nanohertz GWs later, and inhomogeneous remaining cold matter may collapse into (primordial) compact stars [99], which may constitute dark matter if the results are naturally extended to dark sectors [100]. Ejected matter is likely to form halos around central compact stars or disperse, forming structures with more matter surrounding the central core. Besides matter ejection and stochastic nanohertz GWs later than BBN, secondary signals such as electromagnetic and gravitational radiation from accretion might also be observed in the late universe.

Though uniform empirical formulas fitting $S_3/T$ or $\beta/H$ can make calculations more convenient, especially when the numerical error is amplified near the CNPs, it is expected that generally physical quantities, including $\eta$, $\frac{S_3}{T}$, and $\frac{\beta}{H}$, can be determined solely by the effective grand potential $\Omega = -p$ and necessary external parameters such as chemical potential $\mu$, magnetic field, angular velocity, and so on. For dense systems, phase transitions are driven mainly by enthalpy. Numerically, when $c \ll 180$, the difference in entropy density between the true and false vacuum $\Delta s = \frac{\partial \rho_{\text{vac}}}{\partial T} < 0$, which measures the cooling rate, is approximately invariant at small supercooling, i.e., $\rho_{\text{vac}}$ approximately linearly grows from 0 to the value at $T_n$. Detailed numerical relationships and physical understanding are left for future work.

Furthermore, as discussed, to obtain $\beta/H \sim 10$ in the dense system, the potential barrier should be appropriate so that the nucleation temperature $T_n$ is about half of the critical temperature $T_c$. But the extremely precise chemical potential $\mu$ is not that reasonable and can easily be violated by fluctuations. Other QCD first-order phase transitions with magnetic fields or angular velocity are also worth exploring, e.g., a first-order chiral phase transition accompanied by a deconfinement phase transition in the PNJL model with a magnetic field [101], and a first-order topological phase transition with a magnetic field in a chirality imbalanced system [15,16]. It is also worth mentioning that when considering diquark condensation, i.e., the color superconducting phase at high baryon density [102–104], and the coupling between diquarks and quark-antiquarks, the first-order phase transition line at high $\mu$ and low temperature might change to a crossover at very low temperature, but may not change the CNP, and hence the PQNs would still possibly survive.

## ACKNOWLEDGMENTS

We thank F. Gao and J. Schaffner-Bielich for helpful discussions. This work is supported in part by the National Natural Science Foundation of China (NSFC) Grants No. 12235016 and No. 12221005, the Strategic Priority Research Program of Chinese Academy of Sciences under Grant No. XDB34030000.

## DATA AVAILABILITY

No data were created or analyzed in this study.